# WHEN MAJORITY VOTING FAILS: COMPARING QUALITY ASSURANCE METHODS FOR NOISY HUMAN COMPUTATION ENVIRONMENT

Yu-An Sun

Xerox Innovation Group
800 Phillips Road
Webster, NY, 14580, USA
yuan.sun@xerox.com

Christopher Dance

Xerox Innovation Group
6 chemin de Maupertuis
38240 Meylan, France
chris.dance@xrce.xerox.com

**ABSTRACT**

Quality assurance remains a key topic in human computation research. Prior work indicates that majority voting is effective for low difficulty tasks, but has limitations for harder tasks. This paper explores two methods of addressing this problem: tournament selection and elimination selection, which exploit 2-, 3- and 4-way comparisons between different answers to human computation tasks. Our experimental results and statistical analyses show that both methods produce the correct answer in noisy human computation environment more often than majority voting. Furthermore, we find that the use of 4-way comparisons can significantly reduce the cost of quality assurance relative to the use of 2-way comparisons.

## INTRODUCTION

Human computation is a growing research field that holds promise of humans and computers working seamlessly together to implement powerful systems. Algorithmically aggregating outputs from human computation workers is the key to such an integrated human-computer system (Little & Sun 2011). The nature of a human computation system is for workers to self-select tasks to work on, thus results from such an open call are generally noisy with different levels of correctness and quality. Redundancy with majority voting (Bernstein et al. 2010) or independent output agreement (von Ahn & Dabbish 2004) is commonly adopted to address this issue. However, Sun *et al* (2011) and Law & von Ahn (2009) both identified that high quality results reside in a minority of the responses and are often not identified by majority voting. As pointed out in (Law & von Ahn 2011), the limitations of majority voting include 1) the workers can agree on an incorrect answer by chance, 2) workers may have different specialized skills, and 3) the difficulty of the task affects the quality of the responses.

We propose two methods for selecting the best answer in human computation that are based on multi-way comparisons: tournament selection and elimination selection. We conduct proof-of-concept experiments using CrowdFlower.com. Experimental results show that majority voting often produces incorrect answers in situations where the selection methods identify the correct answer. We simulate these methods to benchmark their time complexity, error rates and the costs associated with their deployment, in terms of number of comparisons required, and compare them with selection based on Condorcet voting (Stern 1993).

The main points of this paper are:
1) Both tournament selection and elimination selection produce the correct answer where majority voting fails. However elimination selection typically has a lower error rate when the same number of comparisons is made.
2) With the same cost, 4-way comparison selection schemes have a smaller error rate than pair-wise and 3-way selection schemes.
3) While a Bradley-Terry model (Bradley & Terry 1952) fits our experimental results on $n$-way comparisons for each given $n$, it is not possible to simultaneously describe 2-, 3- and 4-way comparisons with a single joint Bradley-Terry model.

## RELATED WORK

***Quality Control in Human Computation***
Independent agreement and filtering are the two most commonly used quality control methods for human computation. Independent agreement aims to select the best output by majority voting. In the ESP game, the agreement mechanism can be input agreement or output agreement (Law & von Ahn 2009). An output agreement system only accepts image labels agreed by two independent players, and no communication is allowed between them. An input agreement system gives two players a set of inputs to generate an output that they both agreed on, and communication is allowed. We consider only output agreement mechanism as independent agreement and equivalent to majority voting. Games with a purpose that adopt output agreement include the ESP game (von Ahn & Dabbish 2004), HerdIt (Barrington et al. 2009), and Categorilla (Vickrey et al. 2008).

Filtering bad output based on gold questions is another technique for quality control (Le et al. 2010, Oleson et al. 2011). The general idea is that if one can answer the gold question correctly, one also has a higher probability of correctly completing a task. Even though the effectiveness of gold questions has been demonstrated, generating a good set of gold questions remains hard.

*Statistical Modeling for Noisy Pair Comparison*
Pairwise and multi-way comparisons are commonly used for psychology experiments and image quality assessment in conjunction with statistical models such as the multinomial logit (Ben-Akiva & Lerman 1985), Bradley-Terry (Bradley & Terry 1952) and Plackett-Luce (Marden 1995, Ben-Akiva & Lerman 1985) models. Such work typically focuses on estimating the qualities of different items when the pairs to be compared are specified externally. In contrast, we wish to find an efficient algorithm for choosing which pairs to compare in order to select the best answer, as in Adler *et al* (1994). In other words, assuming there is a ground truth best answer, we are trying to determine which observations to make, whereas such statistical models are typically used to analyze the data after the observations have been made.

## PROBLEM DEFINITION

Given $m$ options, our goal is to find the option with the highest quality. However, this task must be accomplished economically in terms of the number of 2-, 3- or 4-way comparisons. Thus, ideally we would like an algorithm which maximizes the probability that the highest quality item is selected subject to a constraint that limits the maximum number of comparisons.

## OUR METHODS

We now present three methods for selecting a best answer in human computation tasks that are based on the hypothesis that humans are better at comparing results to pick the correct one than at producing correct results. All three methods are readily parallelized (at the cost of only a few additional comparisons in the case of elimination selection). This is important as human computation systems are intrinsically parallel in nature.

*Method A: Tournament Selection* This method includes the following steps:

**Input**: $m$ items to be compared using $n$-way comparisons; a pool size $P$ and a fraction $f$
1. Let the initial pool consist of items $[1, 2, ..., m]$.
2. Randomly draw $n$ items from the current pool. Ask one person to select the best of these $n$ items. The selected item goes into a pool of "next generation" answers.
3. Repeat step *2* for $P$ times to generate a new pool of size $P$.
4. *Stopping Condition*: repeat steps 2-3 until some item occupies at least a fraction $f$ of the current pool.
5. The majority item from the current pool is identified as the best item.

In all the experiments in this paper, we set $P = 30$ and $f \in [0.4, 0.9]$. The sampling in step 2 is done uniformly and with replacement, and if there are only $n'$ distinct items sampled out of $n$, then we do a $n'$-way comparison.

The complexity of tournament selection has previously been investigated in (Goldberg & Deb 1991, Fermandez et al. 2010). For our implementation, the number of $n$-way comparisons is $PR$ where $P$ is the size of the pool and $R$ is the number of rounds until the stopping condition is met. The number of rounds $R$ varies with the difficulty of the problem. This gives the method some opportunity to adapt: $R$ typically increases with the number of items to compare and with the difficulty of the problem.

*Method B: Elimination Selection* This method is as follows:

**Input**: $m$ items to be compared using $n$-way comparisons; a number of losses $T$
1. Let the initial pool consist of items $[1, 2, ..., m]$.
2. Randomly draw $n$ distinct items from the current pool. Ask one person to select the best of these $n$ items. Each item other than the best is recorded as receiving one loss.
3. If an item $i$ loses $T$ times, then eliminate item $i$ from the current pool, provided that this leaves a pool of size at least $n$.
4. Repeat steps 2-3 until at most one item has lost fewer than $T$ times.
5. Any item with the minimum number of losses is identified as the best item.

In all the experiments in this paper, we set $T \in [10, 80]$. The sampling in step 2 is done without replacement and so as to equalize the number of times that each item is involved in a comparison.

Elimination selection always makes fewer than $\left(\frac{m}{n-1} + 2\right)T$ $n$-way comparisons, where $m$ is the number of items to compare. In the Appendix we analyze the performance of $n$-way elimination selection and provide a general bound on its error rate. This gives insight into the choice of parameters $T$ and $n$ and into what makes it difficult to select the best option.

*Method C: Condorcet Voting* This method is included as the simplest possible baseline, and is as follows:

**Input**: $m$ items to be compared by $n$-way comparisons and a number of comparisons $k$
1. For each possible $n$-way subset $S$ of the $m$ items, make $k$ comparisons of subset $S$.
2. Identify any item which won the most comparisons as the best item.

Condorcet voting always makes $k\binom{m}{n}$ comparisons, where $\binom{m}{n}$ denotes the binomial coefficient. This may be prohibitive for large $m$ or $n$. It might be anticipated that Condorcet voting will result in a higher error rate than tournament selection and elimination selection for a given number of comparisons, since these other methods avoid making comparisons between items that have already been observed to perform poorly.

**RESULTS**

We first present results on a Chinese idiom translation task that demonstrate that tournament selection and elimination selection can succeed where majority voting fails. We then present a statistical analysis of human computation on five tasks. This enables us to assess the costs and error rates of our methods in real-world settings, and to address the question of whether it is better to use 2-, 3- or 4-way comparisons.

**CHINESE IDIOM TRANSLATION**

*Majority Voting*
We ran experiments to translate the five Chinese idioms listed in Figure 1. This is a challenging task since the literal meanings of idioms are different from their true interpretations. We performed these experiments on CrowdFlower which facilitates the completion of online micro-tasks among a number of labor channels including Amazon Mechanical Turk.

| Chinese Idiom | Groundtruth Translation |
|---|---|
| 一石二鳥 | Achieve two things at the same time |
| 畫虎不成反類犬 | Attempting something beyond one's ability and fail |
| 雞飛狗跳 | Being in chaotic situation |
| 裝聾作啞 | Pretend to be ignorant |
| 三人成虎 | Repeated rumors make people think it's a fact |

*Figure 1: Chinese idioms with corresponding ground truth*

Specific instructions were given to capture true interpretations in English and not to produce literal translations. One unit of human computation task is to translate one idiom and each idiom was translated 30 times (*i.e.* by 30 different workers). We deliberately kept redundancy high to illustrate the large proportion of wrong answers in the crowd's response. Each translation task was paid for $0.03. No further automatic or manual validation was done for the results.

Figure 2 illustrates the number of correct and wrong responses compared to the groundtruth. The error rate for each Chinese idiom translation is above 85% all five of the idioms. In Figure 3, we show the results of majority voting and the percentage of majority votes. Clearly, the majority votes never agree with the ground-truth translation results.

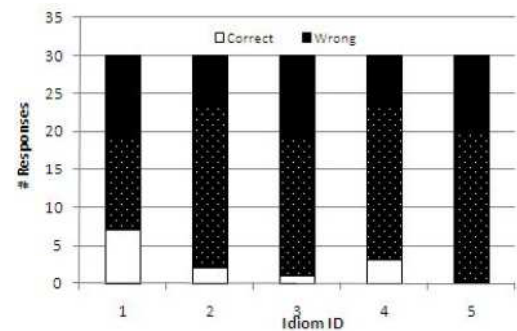

*Figure 2: Number of correct and wrong responses for Chinese idioms translation*

| Chinese Idioms | GroudTruth Translation | Majority Votes Translation | Majority % |
|---|---|---|---|
| 一石二鳥 | Achieve two things at the same time | Two birds with one stone | 43.33% |
| 畫虎不成反類犬 | Attempting something beyond one's ability and fail | Like a dog fails to draw a tiger | 70% |
| 雞飛狗跳 | Being in chaotic situation | Deserted | 60% |
| 裝聾作啞 | Pretend to be ignorant | Dumb | 66.67% |
| 三人成虎 | Repeated rumors make people think it's a fact | Three into a tiger | 66.67% |

*Figure 3: Majority Voting Results*

*Tournament Selection Experiments*

We ran tournament selection (Method A) with pair-wise comparisons on the second Chinese idiom translation. We started with the distinct outputs generated by the CrowdFlower workers. These outputs were as follows, with the 4[th] entry being the correct translation:

1. Like a dog fails to draw a tiger
2. Who are you?
3. None
4. **Attempting something beyond one's ability and fail**
5. Painted tiger anti-dog
6. To try to draw a tiger and end up with the likeness of a dog.

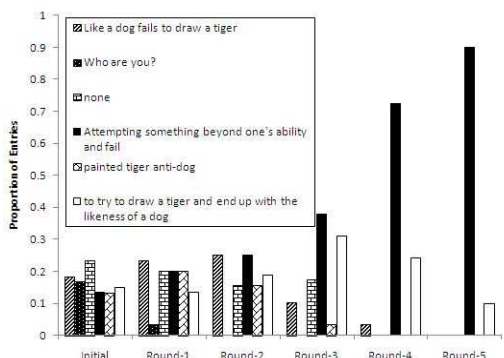

*Figure 4: Tournament selection for one Chinese idiom*

At the end of each round of tournament selection, the proportion of each of the above entries was computed and the corresponding plot is shown in Figure 4. The correct translation was a minority to start off with but it gradually surpassed all other candidates and emerged as a clear winner within five rounds.

Thus it appears that tournament selection is well-suited to such translation tasks since workers find it easier to select the right translation from a list, than to produce the right translation.

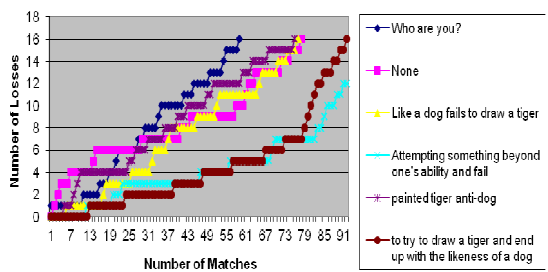

*Figure 5: Elimination selection for one Chinese idiom*

*Elimination Selection Experiments*

We also ran elimination selection (Method B) with pair-wise comparisons on Chinese idiom translation. We use the same set of distinct outputs as in our experiments on tournament selection.

In this experiment, once a translation reaches $T = 16$ losses, it is eliminated from the pool. The final answer is produced when the pool has only one option left. Figure 5 shows the number of losses for each translation versus the number of matches. The correct translation, option 4, is chosen as the final answer after the 91[st] match.

## MODELLING $n$-WAY COMPARISON DATA

In order to predict the costs and error rates of our methods in real-world settings, and to address the question of whether it is better to use 2-, 3- or 4-way comparisons, we need to make some statistical assumptions. Therefore, we now fit several statistical models to a set of 1273 $n$-way comparisons on four tasks, consisting of one Chinese idiom translation and the following three puzzles:

*Puzzle 1*: If you had an infinite supply of water and a 5 quart and 3 quart pail, how would you measure exactly 4 quarts?

*Puzzle 2*: You have a bucket of jelly beans. Some are red, some are blue, and some are green. With your eyes closed, pick out 2 of a like color. How many do you have to grab to be sure you have 2 of the same?

*Puzzle 3*: A chicken and a half can lay an egg and a half in a day in a half. How long will it take for two chickens to lay 32 eggs?

The puzzle data was obtained as follows. We conducted 30 experiments for each puzzle to generate a list of likely answers. A list of six options was compiled including the correct answer and the 5 most common answers produced by human computation workers in these experiments.

To model this data, we make use of the Bradley-Terry model [16], which is as follows. Given a $n$-way comparison between a set of items $K \in \mathcal{K}_n$ where $\mathcal{K}_n$ is the set of all subsets of $\{1, 2, \ldots, m\}$ of size $n$, the likelihood that item $i \in K$ is selected is

$$p(i|\boldsymbol{\gamma}, K) = \frac{\gamma_i}{\sum_{k \in K} \gamma_k}$$

where parameters $\boldsymbol{\gamma} \coloneqq (\gamma_1, \gamma_2, \ldots, \gamma_m)$ with $\gamma_j \geq 0$, describe the quality of the items (the item with the largest $\gamma_i$ is the preferred item).

We test whether there is a preference structure to the data, whether the data corresponds to random clicking by the workers and whether a single set of parameters can predict both $n_1$- and $n_2$-way comparisons for $n_1 \neq n_2$. These possibilities correspond to the following hypotheses, for each task and for each $n$:

$H_{BT}$: *The outcomes of $n$-way comparisons are given by a Bradley-Terry model;*

$H_{SAT}$: *The outcomes of $n$-way comparisons are given by a saturated model (in which the probability that item $i$ wins a comparison $c$ is unrelated to the probability that item $i'$ wins a comparison $c'$ for any $c \neq c'$);*

$H_{MN}$ : *The options selected by a worker are multinomially distributed, independent of which comparison is being made;*

and the following hypothesis for each task:

$H_{JOINT}$ : *The outcomes of 2-, 3- and 4-way comparisons are given by a single joint Bradley-Terry model.*

| Task | | $H_{BT}$ vs $H_{MN}$ | | $H_{SAT}$ vs $H_{BT}$ | |
|---|---|---|---|---|---|
| | N | D | p | D | p |
| translation 2-way | 155 | 65.1 | 0.00 | 2.7 | 0.987 |
| translation 3-way | 135 | 64.2 | 0.00 | 36.1 | 0.417 |
| translation 4-way | 158 | 110.0 | 0.00 | 34.6 | 0.711 |
| puzzle1 2-way | 85 | 1.7 | **0.79** | 10.8 | 0.373 |
| puzzle1 3-way | 108 | 35.3 | 0.00 | 48.9 | 0.060 |
| puzzle1 4-way | 84 | 14.6 | 0.00 | 39.8 | 0.477 |
| puzzle2 2-way | 81 | 11.7 | 0.02 | 4.9 | 0.901 |
| puzzle2 3-way | 107 | 13.6 | 0.00 | 32.3 | 0.597 |
| puzzle2 4-way | 78 | 2.6 | **0.28** | 39.5 | 0.493 |
| puzzle3 2-way | 78 | 10.2 | 0.04 | 4.2 | 0.937 |
| puzzle3 3-way | 114 | 17.8 | 0.00 | 53.1 | **0.026** |
| puzzle3 4-way | 90 | 28.3 | 0.00 | 38.5 | 0.537 |

*Table 1: Likelihood-ratio tests of hypotheses for four tasks (symbols described in text)*

Results for these hypothesis tests are given in Tables 1 and 2. In these tables, N is the number of comparisons in the data; the deviance is $D = -2(L_C - L_S)$ where $L_C$ and $L_S$ are the log-likelihoods under the complex and simple hypotheses respectively; and the p-value is obtained by the $\chi^2$-approximation. In all but one case, the data has a clear preference structure ($H_{SAT}$ vs $H_{BT}$). In all but two cases, random clicking can be discarded as a hypothesis ($H_{BT}$ vs $H_{MN}$), and in only one case might a single set of parameters predict $n$-way comparisons for different $n$ ($H_{JOINT}$ vs $H_{BT}$).

It is also interesting to see how often the population of workers selected the right answer. Given that the Bradley-Terry hypothesis is plausible, Table 3 lists the maximum likelihood estimates of the parameters $\gamma_i$ for the known best item and for the *competing item* (*i.e.* that with the next largest merit if the known best option does not win or the option with the largest estimated merit if the known best option loses). The table shows that the workers tend to prefer the right answer in only 5 out of 12 experiments!

| Task | | $H_{JOINT}$ vs $H_{BT}$ | |
|---|---|---|---|
| | N | D | p |
| translation | 448 | 29.2 | 0.00 |
| puzzle1 | 277 | 45.8 | 0.00 |
| puzzle2 | 266 | 25.8 | 0.00 |
| puzzle3 | 282 | 15.8 | **0.11** |

*Table 2: Test for a single joint Bradley-Terry model (symbols described in text)*

| Task | 2-way | 3-way | 4-way |
|---|---|---|---|
| translation | 0.478 | 0.263 | 0.536 |
| | 0.285 | 0.335 | 0.280 |
| puzzle1 | 0.104 | 0.430 | 0.110 |
| | 0.210 | 0.256 | 0.387 |
| Puzzle2 | 0.253 | 0.218 | 0.290 |
| | 0.236 | 0.252 | 0.228 |
| Puzzle3 | 0.153 | 0.223 | 0.153 |
| | 0.383 | 0.337 | 0.478 |

*Table 3: Estimates of parameters $\gamma_i$ normalized so that $\sum_{i=1}^{6} \gamma_i = 1$. Top row for each task: known best option; second row: competing option (defined in text)*

## COMPARISON OF SELECTION METHODS

Our bound on the error rate of elimination selection (see Appendix) describes its performance for a rather general family of comparison probabilities. However, we would like to understand the error rate and mean number of games that might result when applying both tournament and elimination selection on data observed in real experiments. In this section, we take a Bayesian approach to those questions.

If the data had an exact Bradley-Terry distribution, in which the known best item had the highest winning probability, the algorithms would perform as shown in Figure 6. In this plot, the error rate converges to zero as the number of comparisons increases and there is a clear performance improvement as we move from 2- to 3- to 4-way comparisons. However, the real-world data might not have an exact Bradley-Terry distribution and perhaps the actual distribution does not satisfy the assumptions under which elimination selection converges. Even if the data did have an exact Bradley-Terry distribution, we do not know the parameters of that distribution. Furthermore, the known best item might not be preferred by the workers.

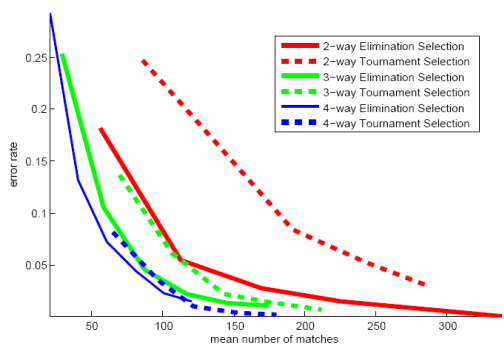

*Figure 6: Error rates and mean number of $n$-way comparisons for choices following an exact Bradley-Terry distribution.*

Instead, we shall aim is to estimate the posterior expected error rate and the mean number of comparisons made by each algorithm, given the $n$-way comparison data analyzed in the previous section. We do so via a Monte Carlo Markov Chain method. Based on our findings in the previous section, we assume that the selection probabilities are either drawn from a Bradley-Terry distribution with a uniform prior on $[0,1]^m$ for the parameters $\gamma_i$, or are drawn from the saturated model with a uniform Dirichlet prior on the probabilities that item $i$ wins a comparison with items $j, k, l$.

In detail our method is as follows, we draw 1000 samples of the winning probabilities from the posterior using 5000 accepted steps of Metropolis-Hastings, where the first 1000 steps were treated as burn-in and the remaining steps were thinned by extracting only one in every 4 samples. To check that this was a sufficient number of iterations, we used the R-package CODA (Plummer et al. 2006). For each such sample we ran each selection algorithm and the error rate was estimated as the fraction of samples for which the selection algorithm did not return the item with the highest sampled merit.

Figures 7 and 8 show the results of this MCMC method when applied to the Chinese idiom translation task assuming the saturated model and the Bradley-Terry model. For tournament selection we varied the fraction $f$ in the stopping criterion from 0.5 to 0.9 and for elimination selection we varied stopping parameter $T$ from 10 to 80. In Figure 9, the balanced experiment is equivalent to the Condorcet voting since every permutation of comparison is included.

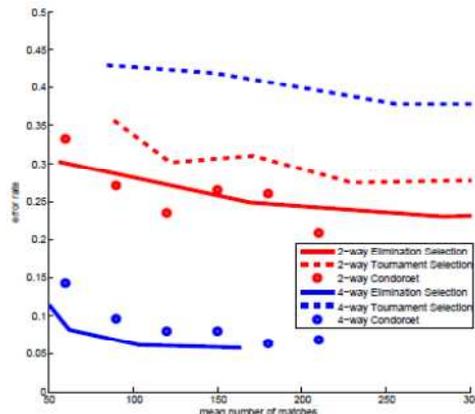

*Figure 7: Error rates and mean number of comparisons for preferences sampled from the posterior of data from the Chinese idiom translation task assuming the saturated model.*

In neither plot does the error rate converge to zero, as there is always a non-zero probability that the population does not prefer the known best option. Indeed, the saturated model applied to 4-way tournament selection produces a rather high error rate. This is because 4-way tournament selection sometimes makes 3- or 2-way comparisons (e.g. if items {1, 1, 2, 3} are sampled), and the 3-way data strongly supports a model in which the known best item is not preferred.

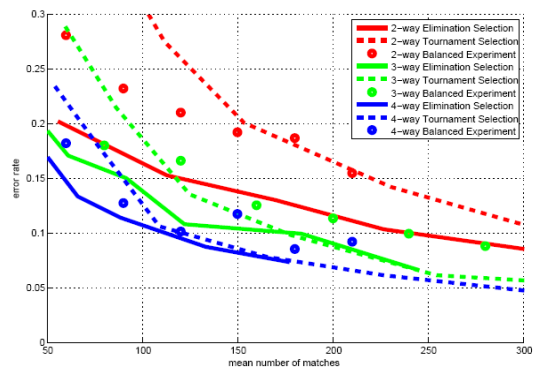

*Figure 8: Posterior error rates and mean number of comparisons given data from the Chinese idiom translation task assuming a Bradley-Terry model.*

All plots in this section clearly illustrate the advantage of the elimination selection method over

Condorcet voting and over the tournament selection method for 2-way comparisons. However, the gap between the methods decreases for both 3-way and 4-way comparison cases (with the exception of tournament selection in the saturated model). Furthermore, there is a clear benefit in moving from 2- to 3- to 4-way comparisons. Comparable plots are obtained in the case of the four puzzles in those cases where the data supports the hypothesis that the known best answer is preferred by the population.

**CONCLUSIONS**

We proposed two methods for selecting the best answer in difficult human computation tasks: tournament selection and elimination selection. Both methods successfully produce the correct answer where majority voting fails by conducting proof-of-concept experiments. We conducted a statistical analysis of multi-way comparisons obtained from five human computation tasks. While a Bradley-Terry model fits the data on $n$-way comparisons for each given $n$, it is not possible to simultaneously describe 2-, 3- and 4-way comparisons with a single joint Bradley-Terry model.

Using a Bayesian approach based on this analysis, we compared the cost and error rate of each method with Condorcet voting. This comparison showed that for the same cost, 4-way comparison selection schemes give a smaller error rate than pair-wise selection schemes.

**FUTURE WORK**

Our work demonstrated applicability of tournament and elimination selection methods as quality assurance on human computation systems. Future work includes exploring a mixture model for Bradley-Terry to better capture workers' intent, a hybrid method of both selection and filtering, and different type of human computation tasks.

## APPENDIX: HOW OFTEN DOES $n$-WAY ELIMINATION SELECTION MAKE ERRORS?

We analyze the performance of the $n$-way elimination selection method. Such an analysis was sketched in (Adler et al. 1994). We complete that sketch and extend it to the case of $n$-way comparisons. Thus, this appendix justifies the use of $n$-way elimination selection and gives insight into the selection of parameter $T$.

Firstly, we must ensure that the probability that an item wins a match corresponds to having some best item, which will be identified as item 1. To do so we make a *discriminating assumption,* which is a generalization of the assumption made in [? Adler]. Say item $j \neq 1$ and item 1 are both matched with some set $K$ of items to be compared. For instance, if $n = 3$ and $K = \{2,3\}$ then we consider two matches involving items $1,2,3$ and items $j,2,3$, whereas if $K = \{1,2,j\}$ then we consider only one match involving items $1, 2$ and $j$. Let $X = 1$ (or 0) if item 1 loses (wins) its match and let $Y = 1$ (or 0) if item $j$ loses (wins) its match. We assume that for some $\delta > 0$, for all items $j \neq 1$, for all sets of items $K$ (as above),
$$P(X = 0, Y = 1 \mid X + Y > 0, K)$$
$$\geq P(X = 1, Y = 0 \mid X + Y > 0, K) + \delta.$$
That is, the probability that only item $j$ loses is at least $\delta$ larger than the probability that only item 1 loses, given that at least one item loses. Provided that there is some best item, it is straightforward to verify that the discriminating assumption is satisfied for many widely-used multi-way comparison models, such as the Plackett-Luce model (Ben-Akiva & Lerman 1985).

We are now ready to state our main result, which relates the probability $\epsilon$ that $n$-way elimination selection does not select item 1, to the choice of parameter $T$ and to the difficulty of the selection problem, as represented by $\delta$.

**Proposition A1.** *Suppose that $n$-way elimination selection is run for $m \in \mathbb{N}$ items with parameter $T \in \mathbb{N}$ and that the loss probabilities satisfy the discriminating assumption with parameter $\delta$. Then the failure probability $\epsilon$ is bounded by*
$$\epsilon \leq m \exp(-\delta^2 T/4).$$

**Proof.** We consider any item $j \neq 1$ and show that the probability that item 1 is eliminated before item $j$ is at most $\exp(-T\delta^2/4)$, using Azuma's inequality. The proposition then follows directly from the union bound.

First, let $A_t$ count the losses of item 1 and $B_t$ count the losses of item 2 at the subsequence of rounds $t$ where either 1 or $j$ or both lose a match. The probability that 1 is eliminated before $j$ is

$P(A_\sigma \geq B_\sigma)$ where $\sigma := \inf\{t \mid max\{A_t, B_t\} = T\}$.
It turns out to be easier to consider a different stopping time $\tau$. In particular, we let $\tau$ be the first time that $A_t + B_t \geq 2T$, where we imagine that items 1 and $j$ continue playing matches rather than being eliminated at time $\sigma$. Since $A_t$ and $B_t$ are non-decreasing and $\tau \geq \sigma$ for any realisation, it follows that $A_\sigma \geq B_\sigma$ implies that $A_\tau \geq B_\tau$. Hence

$$P(A_\sigma \geq B_\sigma) \leq P(A_\tau \geq B_\tau).$$

Now we will define a martingale difference sequence $M_t$ to which we can apply Azuma's inequality. Let $F_t$ denote the history of games up to and including time $t$ and let $X_t := A_t - A_{t-1} - (B_t - B_{t-1})$. Define the process $Z_t := \max\{-c_t, X_t\}$ where

$$c_t := (P(X_t = 1 \mid F_{t-1}) + \delta)/P(X_t = -1 \mid F_{t-1}).$$

By the discriminative assumption $c_t \leq 1$, hence
$$\sum_{s=1}^{t} Z_s \geq A_t - B_t$$
for any realization. Furthermore $E[Z_t \mid F_{t-1}] = -\delta$, thus the martingale difference sequence
$$M_t := t\delta + \sum_{s=1}^{t} Z_s$$
satisfies
$$P(A_\tau \geq B_\tau) \leq P(M_\tau \geq \tau\delta) \leq P(M_\tau \geq T\delta)$$
where the second inequality follows since $\tau \geq T$ for all realizations.

Now consider the stopped martingale difference sequence $M_{\min\{t,\tau\}}$. Since $\tau \leq 2T$ for any realization we have
$$P(M_\tau \geq T\delta) = P\big(M_{\min\{2T,\tau\}} \geq T\delta\big).$$
We now apply Azuma's inequality, noting that martingale $M_{\min\{t,\tau\}}$ has differences of range at most two, since $\sup\{M_{t+1} - M_t\} - \inf\{M_{t+1} - M_t\} \leq 2$, giving
$$P(M_\tau \geq T\delta) \leq \exp\left(-\frac{(T\delta)^2}{2 \cdot 2T}\right) = \exp(-T\delta^2/4).$$

Thus item 1 is eliminated before item $j$ with probability at most $\exp(-T\delta^2/4)$.

The proposition now follows directly from the union bound applied for all possible items $j \neq 1$. ∎